\documentclass[published]{JHEP3} 
\JHEP{00(2002)000}

\JHEPspecialurl{http://jhep.sissa.it/JOURNAL/JHEP3.tar.gz}

\usepackage{epsfig,multicol,bbm}

\newcommand\fverb{\setbox\pippobox=\hbox\bgroup\verb}
\newcommand\fverbdo{\egroup\medskip\noindent%
			\fbox{\unhbox\pippobox}\ }
\newcommand\fverbit{\egroup\item[\fbox{\unhbox\pippobox}]}
\newbox\pippobox

\title{Deciphering the properties of the medium produced in heavy ion
collisions at RHIC by a pQCD analysis of quenched large $p_{\perp}$ 
 $\pi^0$ spectra}

\author{Rudolf Baier$^{\rm 1}$~and Dominique Schiff$^{\rm 2}$\thanks{Unit\'e Mixte de Recherche - CNRS -UMR 8627}\\
$^{\rm 1}$ Physics Department, University of Bielefeld, D-33501 Bielefeld, Germany\\
	E-mail: \email{baier@physik.uni-bielefeld.de}\\
	$^{\rm 2}$ LPT, $B\hat{a}t.~210$, Universit\'e Paris-Sud, F-91405 Orsay, France\\
	 E-mail: \email{Dominique.Schiff@th.u-psud.fr}}
\received{} 		
\revised{}
\accepted{}		

\preprint{\hepth{9912999}}	

\abstract{We discuss the question of the relevance of perturbative QCD
calculations for analyzing the properties of the dense medium produced
in heavy ion collisions.
Up to now leading order perturbative estimates have been worked out
and confronted with data for quenched large $p_{\perp}$ hadron spectra.
Some of them are giving paradoxical results, contradicting the
perturbative framework and leading to speculations such as the formation
of a strongly interacting quark-gluon plasma.
Trying to bypass some drawbacks of these leading order analysis 
and without performing detailed
 numerical investigations, we collect evidence in favour of a consistent
description of quenching and  of the  characteristics
of the produced medium within the pQCD framework.}


\begin{document} 

\section{Introduction}
There is convincing evidence that hadron spectra at high $p_{\perp}$
in nuclear collisions are strongly suppressed
\cite{RHIC,Muller:2005wi}.
For neutral pion production 
 in Au-Au collisions at RHIC the nuclear modification factor $R_{AA}(p_{\perp})$
is measured to be about $0.2 - 0.3$ in the range of transverse momenta
as large as $10~GeV/c < p_{\perp} < 20~GeV/c$
\cite{Akiba:2005bs,Isobe:2005mh,Tannenbaum:2006ch}.
This quenching is usually  attributed to radiative parton energy loss,
 for which the relevant expressions
are obtained from perturbative QCD calculations
\cite{Baier:2000mf,Kovner:2003zj}. They are used to extract and analyze the
properties of the deconfined medium produced in the collisions. As a consistency
 requirement these properties should be compatible with the perturbative
 framework.

However, the detailed analysis given in \cite{Eskola:2004cr,Salgado:2005zp}
 results in
a (time-averaged) value for the transport coefficient $\hat{q}$ which
characterizes the medium, exceeding $5~ GeV^2/fm$, almost a factor of 
$10~$ (or even more) bigger than a typical perturbative estimate at
 the energy density
expected for $\sqrt{s_{NN}} = 200~GeV$ Au-Au collisions! 

An independent work \cite{Arleo:2006xb},
based on \cite{Arleo:2002kh}, calculating the quenching factor for
hard pion production and comparing it with data measured at RHIC
obeys this requirement of perturbative consistency: an averaged value
of $\hat{q}\simeq 0.3 - 0.4 ~GeV^2/fm$ is found, corresponding to an energy density
$\epsilon \simeq 2 ~GeV/fm^3$, as expected from pQCD for a deconfined
equilibrated plasma. However, this value is based on imposing arbitrarily
a mean path length for the jet of about $L= 5~fm$,
whereas for the denser medium discussed in \cite{Eskola:2004cr} a
characteristic path length of $L \simeq 2~fm$ is obtained.

A related work on the nuclear modification factor for leading large $p_{\perp}$
hadrons (pions) \cite{Turbide:2005fk}, assuming a thermalized medium (and
also taking into account absorption of thermal partons), shows that a value of
$R_{AA}$ is obtained which is compatible with measurements and the perturbative
framework. This work based on the AMY \cite{Arnold:2002ja} formalism describes the
coherent gluon radiation, incorporating the 
Landau-Pomeranchuk-Migdal  (LPM) effect in the range of
gluon energies $\omega > \omega_{BH}$, where $\omega_{BH}$ 
(sometimes denoted by $E_{LPM}$) corresponds to the
transition energy to the incoherent
Bethe-Heitler radiation regime.

The analysis in \cite{Eskola:2004cr}, following the explicit calculations of 
"quenching weights" \cite{Salgado:2003gb}, does not distinguish between these
two regimes. On the other hand, it does impose a kinematical constraint, taking 
properly into account the effect due to the transverse momentum phase space of
the emitted gluon. This effectively constrains the soft LPM emission by
imposing a lower energy cut-off $\hat{\omega}$, which depletes the gluon
energy distribution.

In this note we argue that the introduction of this cut-off $\hat{\omega}$
is a priori not accurate enough because, as we shall see below,
$\hat{\omega}$ is (much) smaller than $\omega_{BH}$ and thus not relevant for
the LPM regime. This actually is one crucial reason for the large value
of $\hat{q}$ found in \cite{Eskola:2004cr}. We instead show that using
$\omega_{BH}$ as the proper cut-off for the validity of soft LPM gluon emission
may lead to values of  $\hat{q}$ compatible with perturbative estimates.

The line of our arguments treating radiative energy loss
follows the BDMPS \cite{Baier:1996kr,Baier:1998kq} - Zakharov
 \cite{Zakharov:1997uu,Zakharov:2004vm} -
Wiedemann \cite{Wiedemann:2000tf} approach, as it is
reviewed in \cite{Baier:2000mf,Kovner:2003zj}.
The trigger bias induced by the steeply falling large $p_{\perp}$ vacuum
production cross section of produced hadrons/neutral pions is
 treated as described in \cite{Baier:2001yt}.
 
 The infrared sensitivity of the quenching factor has already
 been commented upon
in [20], where it is emphasized that the energy $\omega_{BH}$
plays a central role.

\section{Limits on the quenching factor}

\def\E{\epsilon}

As it is discussed in rather great detail in \cite{Baier:2001yt} the
quenching effect is expressed by the factor
\begin{equation}\label{eq:e1}
Q(p_\perp) \>=\> \int d\E\, D(\E)~ \left(\frac{d\sigma^{{\rm vacuum}}
(p_\perp+\E)/dp_\perp^2}{d\sigma^{{\rm vacuum}}
(p_\perp)/dp_\perp^2}\right)~,
\end{equation}
where it is justified to express the probability $ D(\E)$ for emitting the
energy $\E$ into the medium by a Poissonian energy distribution
\begin{equation}  \label{eq:e2}
D(\E) = \sum^\infty_{n=0} \, \frac{1}{n!} \,
\left[ \prod^n_{i=1} \, \int \, d\omega_i \, \frac{dI(\omega_i)}{d \omega} 
\right] \delta \left(\E - \sum_{i=1}^n  \omega_i\right)
\cdot \exp \left[ - \int d\omega \frac{dI}{d\omega} \right]~. 
\end{equation}
This expression is assumed to be valid for the emission of soft
$\underline{primary}$ gluons. In the LPM regime 
$\omega \ge \omega_{BH}$  the bremsstrahlung spectrum  $dI/d\omega$ 
is given in \cite{Baier:1998kq}.
Correspondingly the multiplicity of LPM gluons with energies larger than
$\omega$ is given by
\begin{equation}\label{eq:e3}
  N\left( {\omega}\right) \equiv \int_\omega^\infty d\omega'\,
 \frac{dI(\omega')}{d\omega'}~.
\end{equation}
For $\omega$ significantly less than the characteristic energy \cite{Salgado:2003gb},
\begin{equation}\label{eq:e4}
\omega_c = \frac{1}{2} \hat{q} L^2
\end{equation}
 but larger than $\omega_{BH}$,
the number of gluons is well approximated by
\begin{equation}
\label{eq:e5}
   N(\omega) \simeq \frac{2 \alpha_s C_R}{\pi}\left[\,  \sqrt\frac{2 \omega_c}{\omega}
 +\ln2\ln{\frac{\omega}{\omega_c}} 
- 1.44 \> \right]~.
\end{equation}

The following remarks are crucial for the subsequent analysis:
\begin{itemize}
\item
The probability $D(\E)$ is normalized by $\int~d\E D(\E) = 1$.

\item
The ratio of cross sections in (\ref{eq:e1}) is well
approximated by
\begin{equation}
\label{eq:e6}
 \frac{d\sigma^{{\rm
vacuum}}(p_\perp+\E)/dp_\perp^2}{d\sigma^{{\rm vacuum}}
(p_\perp)/dp_\perp^2} \>\simeq\>
\left(\frac{p_\perp}{p_\perp+\E}\right)^n
\>\simeq \> \exp\left(-\frac{n\E}{p_\perp}\right)~,
\end{equation}
when expressed in terms of 
an  effective exponent $n$,
which may depend on $p_{\perp}$, $ n = n(p_{\perp})$.
In the following the approximation (\ref{eq:e6}) is used.

\item
Concerning the underlying parton interactions one has to
distinguish quark versus gluon jets. Since our concern is mainly
neutral pion production at RHIC in the range
$10 < p_{\perp} < 20~GeV/c$,  we effectively assume a dominating quark jet,
radiating off (soft) gluons, therefore we take $C_R = C_F = 4/3$
in (\ref{eq:e5}). This assumption is supported by the analysis in
\cite{Eskola:2004cr}.

\item
The quenching factor $Q(p_{\perp})$ corresponds to the experimentally
measured ratio \cite{Tannenbaum:2006ch}
\begin{equation} 
\label{eq:e7}
R_{AA}(p_{\perp}) = \frac{d N_{AA}}{<N_{coll}> dN_{NN}}~,
\end{equation}
for  central A-A collisions versus nucleon-nucleon (NN) collisions. 
In the following only neutral pion
production at  pseudo-rapidity $\eta = 0$ is considered.

\item
The transverse momenta of the produced pions
 are not asymptotically large, although only
leading order pQCD calculations are considered.

\end{itemize}

Neglecting in the following the contribution from  Bethe-Heitler
emission (Appendix A), the quenching factor due to LPM emission becomes
\begin{equation} 
\label{eq:e8}
Q(p_\perp) \simeq \int_0^\infty d\E \,
D(\E)\,\exp{\left\{ -\frac{n \E}{p_\perp} \right\} }
\>=\>\exp\left\{ - \int^{\infty}_{\omega_{BH}}\,\frac{dI}{d\omega} 
\left[1-\exp\left(-\frac{n\omega}{p_{\perp}}\right)\right] 
~d\omega \right\} ~,
\end{equation}
where we take as a lower cut-off the energy $\omega_{BH}$, which indicates
the transition between the Bethe-Heitler and the LPM regime.
By partial integration and with (\ref{eq:e3}), the quenching factor becomes
\begin{equation} 
\label{eq:e9}
Q(p_\perp)= 
\exp\left\{ -N(\omega_{BH})\left[1-\exp\left(-\frac{n\omega_{BH}}{p_{\perp}}
\right)\right] \right\} 
\cdot \exp\left\{ - \frac{n}{p_{\perp}}~\int^{\infty}_{\omega_{BH}}
N(\omega) \exp\left(-\frac{n \omega}{p_\perp}\right) 
{d\omega} \right\}~.
\end{equation}
As $N(\omega)$ decreases with increasing gluon energy \cite{Baier:2001yt},
one finds the following two bounds for $Q(p_{\perp})$:
\begin{equation}
\label{eq:e10}
   Q_{min} (p_{\perp}) \> = \> \exp {\left[ -N(\omega_{BH})\right]}~,
\end{equation}
i.e. by replacing $N(\omega) = N (\omega_{BH})$ in the integrand of
the integral in (\ref{eq:e9}),
and 
\begin{equation}
\label{eq:e11}
   Q_{max}(p_\perp) \>= \>
\exp \left\{ - N(\omega_{BH}) 
\left[1 -\exp\left(-\frac{n\omega_{BH}}{p_{\perp}}\right) \right]
\right\} ~,
\end{equation}
i.e. by neglecting the second  exponential factor of (\ref{eq:e9}).
So that, the experimental ratio is required to be bounded as follows,
\begin{equation}
\label{eq:e12}
 Q_{min}(p_{\perp}) < R_{AA}(p_{\perp}) < Q_{max} (p_{\perp})~.
\end{equation}
One notes that actually $Q_{min}(p_{\perp})$ does not depend on $p_{\perp}$,
and that $Q_{max}(p_{\perp})$ approaches $1$ for large $p_{\perp} >> n\omega_{BH}$.
For a fixed value of $p_{\perp}/n \simeq O(1~GeV)$, the interval
$Q_{max} - Q_{min}$ becomes rather tight, when $\omega_{BH}$ is not a small
energy.

When comparing the quantities $Q_{min,max}$  with the ones defined in
\cite{Eskola:2004cr,Salgado:2003gb}
$Q_{min}$ is related to the
discrete weight of the probability distribution $D(\epsilon)$ by
$ p_0 = Q_{min}$,
where, instead of the soft  cut-off $\hat{\omega}$ the Bethe-Heitler one
$\omega_{BH}$ has to be taken, as required by the LPM radiation
formalism.
According to our analysis $Q_{max}$ is the relevant, important
quantity,
 related to the
continuous part of $D(\epsilon)$, which translates to $Q_{max} - Q_{min}$.

One may notice that the suppression factor
$Q(p_{\perp})$ of (\ref{eq:e9}) does satisfy (for the same value of
$\omega_c$)
\begin{equation}
\label{eq:exq}
Q(p_{\perp}, \omega_{BH}) \ge Q(p_{\perp}, \hat{\omega})~,
\end{equation}
where the value of the lower cut-off is indicated.
But here is where $Q_{max}$ comes into the game.
Its {\underline {actual}} value and the ones of the various
energy scales cannot be left out of the discussion~!

Typical estimates of the medium characteristics as constrained by the
experimental results are discussed in the next section. In particular
we find that $\omega_{BH} \sim 1.5 - 2.0~GeV$. As a consequence,
for values of $p_{\perp}/n \sim 1 ~GeV$, the bounds $Q_{min}, Q_{max}$
are indeed constraining. On the contrary, taking $\hat{\omega}$
as the lower energy cut-off,  one finds that due to $\hat{\omega} << \omega_{BH}$,
$Q_{min}$ and $Q_{max}$ differ significantly and the relation (\ref{eq:e12})
is no longer constraining. 

Replacing $\omega_{BH}$ by $\hat{\omega}$ in the estimate for
$Q_{max}$, the constraints are not very tight indeed:
as an example when choosing $\hat{\omega} \simeq 0.45~ GeV$,
guided by relation (3.6) to be derived in the next section,
  we find $Q_{max} \simeq 0.55$, instead of $Q_{max} \simeq 0.30$.
In both cases $Q_{min} = 0.2$.

\section{Kinematical and consistency constraints}

As mentioned earlier, the detailed discussion given in \cite{Eskola:2004cr}
in order to determine the medium induced gluon radiation intensity $dI/d\omega$
takes into account the kinematical constraint associated to the transverse 
momentum phase space of the emitted gluon. This constraint is not implemented
in earlier works \cite{Baier:1996kr}. The effect of the kinematical
limitation is obtained by estimating the ratio $k_{\perp}/\omega$ in the LPM
regime: in this coherent regime the transverse momentum $k_{\perp}$ of
the emitted gluon may be given by
\begin{equation}
k_{\perp}^2 \simeq \frac{t_{coh}}{\lambda_g}~ \mu^2~,
\label{eq:k1}
\end{equation}
where $\mu$ is the typical transverse momentum transfer in a single scattering
(i.e. the Debye mass screening the gluon exchange) and $\lambda_g$ 
the gluon mean free path, such that $t_{coh}/\lambda_g$ is the
number of coherent scattering centers in the medium which the gluon
encounters before being emitted after the time 
$t_{coh}\simeq 2 \omega/k_{\perp}^2$. One finds
\begin{equation}
k_{\perp}^2 \simeq \sqrt{2 \hat{q} \omega}~,\label{eq:k2}
\end{equation}
where  $\hat{q} \simeq \mu^2/\lambda_g$. As a consequence, since $k_{\perp} \le \omega$,  
gluons have to be emitted dominantly above the energy $\hat{\omega}$
defined by
\begin{equation}
\hat{\omega} = (2 \hat{q})^{1/3} = 
\omega_c {\left(\frac{2}{R}\right)}^{2/3}~,
\label{eq:k3}
\end{equation}
where it is convenient to introduce the dimensionless parameter
\cite{Baier:2001qw,Salgado:2002cd,Salgado:2003gb}
\begin{equation}
R = \omega_c L = \frac{1}{2} \hat{q} L^3~.
\label{eq:k3a}
\end{equation}

Now, we should take into account that the multiple scattering formalism
used throughout the derivation of  $dI/d\omega$ requires the
condition \cite{Gyulassy:1993hr,Baier:1996kr}
\begin{equation}
\lambda_g \mu >> 1 ~.
\label{eq:k4}
\end{equation}
Using the fact that $\omega_{BH}$ may be expressed as
$\omega_{BH} \simeq \lambda_g \mu^2$ we obtain the following
parametric inequality
\begin{equation}
\frac{\hat{\omega}}{\omega_{BH}} \sim 
\frac{2^{1/3}}{(\lambda_g \mu)^{4/3}} << 1 ~.
\label{eq:k5}
\end{equation}
As a consequence of this inequality, the Bethe-Heitler energy remains the
proper and relevant lower limit for the validity of the LPM gluon
emission spectrum.

Let us summarize what is already known about the implementation of the
various above mentioned constraints.
In the BDMPS framework \cite{Baier:1996kr,Baier:1998kq} where no kinematical
cut-off  is imposed on the $k_{\perp}$ integration, i.e. in the limit 
$R \rightarrow \infty$, the gluon number only depends on the ratio
$\omega/\omega_c$: $~N(\omega) = N(\omega/\omega_c)$. As a consequence
the resulting quenching factor $Q(p_{\perp})$ is effectively a 
scaling function in the variable $X= p_{\perp}/(n \omega_c)$ 
\cite{Baier:2001yt}, such that in the relevant analysis given in 
\cite{Arleo:2006xb}
only the characteristic gluon energy $\omega_c$ is extracted
from the  the neutral pion single-inclusive data measured by the PHENIX
Collaboration in Au-Au collisions \cite{Akiba:2005bs,Isobe:2005mh},
and found to be   $\omega_c = 20 - 25~GeV$.
Moreover, as already mentioned, the path length is arbitrarily chosen so that
the medium characteristics are not quantitatively constrained.
In \cite{Eskola:2004cr} the typical value of the parameter $R$
 relevant for the
description of RHIC  data on pion production
at $\sqrt{s_{NN}} = 200~GeV$ is estimated to be $R \simeq 1000$,
 equivalent to $\hat{\omega}
\simeq 0.016~ \omega_c$. 

\section{Comparison with experiment}

We shall first discuss a few semi-quantitative estimates of the parameters 
describing the medium extracted from the comparison with data within the
framework described in the previous sections.

\subsection{Averages}

We concentrate
on the data for $R_{AA}(p_{\perp})$ with $p_{\perp} \ge 10~GeV/c$:
 $~R_{AA}(p_{\perp}) \simeq 0.2$,
remaining essentially flat up to $p_{\perp} \simeq 20~GeV/c$.

In order to start the discussion, we use these data to obtain
values for $Q_{min}$ and $Q_{max}$,
corresponding to the experimental error bars.
Taking $Q_{min} \simeq 0.2$ leads to 
$N(\omega_{BH}/\omega_c) \simeq 1.6$, which allows to estimate
$\omega_{BH}/\omega_c \simeq 3.5 \cdot 10^{-2}$ when using (\ref{eq:e5})
and $\alpha_s = 1/2$ (see  also Fig. 1). When we take $p_{\perp}/n \simeq O(1~GeV)$,
observing that the experimental error bars allow us to fix the value of 
$Q_{max} < 0.3$, we deduce the typical value of $\omega_{BH}$.
From (\ref{eq:e11}) we obtain $\omega_{BH} \ge 1.4~GeV$,
in agreement with thermal estimates (Appendix B).
Taking $\omega_{BH} \simeq 1.6~GeV$, we find $\omega_c \simeq 45 ~ GeV$.
We note that this value is bigger by a factor $2$ than the one extracted
in \cite{Arleo:2006xb}.

Depending on the typical average path length 
we derive estimates for the $\underline{\rm{time-averaged}}$ 
transport coefficient,
\begin{equation}
\hat{q} = \frac{2\omega_c}{L^2} \simeq \frac{18}{(L[fm])^2}
 ~\frac{GeV^2}{fm}~,
\label{eq:c1}
\end{equation}
i.e. $\hat{q} \simeq 1.1~ GeV^2/fm$ for $L = 4~fm$,
and  $\hat{q} \simeq 2.0~ GeV^2/fm$ for $L = 3~fm$. This 
corresponds to values of $R \sim 900~ - ~700$, not far from
 $R \simeq 1000$ in \cite{Eskola:2004cr}.

We observe that these estimates are indeed valid beyond the
$R= \infty$ limit, since in the region of  $\omega_{BH}/{\omega_c} >
3.5~10^{-2}$ the sensitivity on $R$, for $R \ge 1000$, becomes weak.

The cut-off $\omega_{BH}$ is effectively of order $1-2 ~GeV$
  whereas $\hat{\omega}$, as imposed by Eskola et al. [8]  is
also $~ 1~ GeV$, when using relations given in [14].
One may wonder then why  choosing one or the other  is a crucial
feature, independently of judging the validity of using the LPM spectrum
 away from $\omega >
\omega_{BH}$. One way to understand qualitatively this fact
 is that in  the  analysis of Ref.~[14], the
$\hat{\omega}$ cut-off is implemented effectively in the 
$k_{\perp}$  integration for the emitted
gluon whereas here the $\omega_{BH}$ cut-off appears as an IR cut-off
 on the emitted gluon energy in
the expression of $Q(p_{\perp})$. That has consequences on
 the effectiveness of the cut-off.

\FIGURE{\epsfig{file=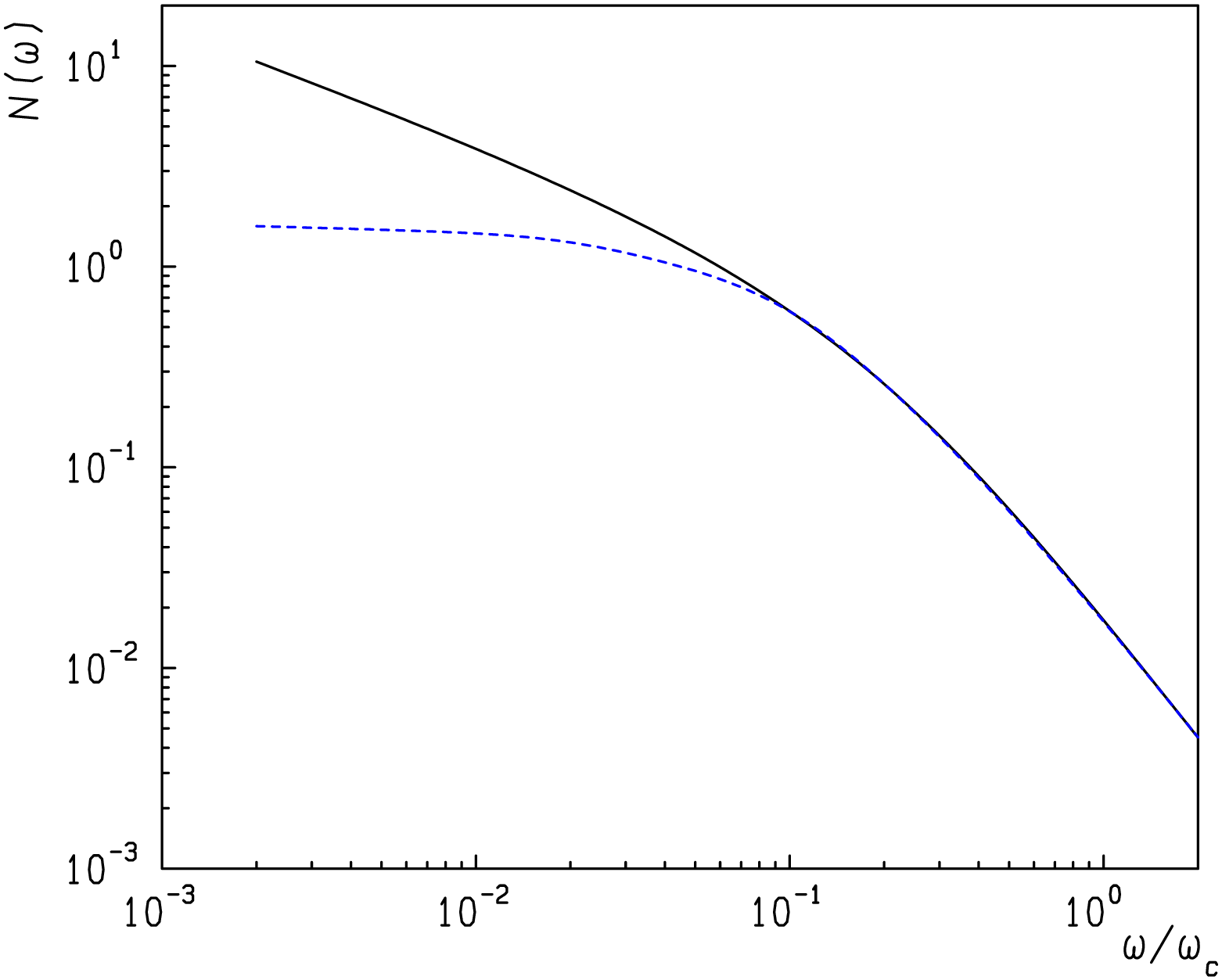, width=100mm}
\caption{ The gluon multiplicity as a function of $\omega$,
for $R = \infty$ (solid curve) and $R= 1000$ (dashed curve) taken
from \cite{Salgado:2003gb}, but for $\alpha_s = 1/2$. }\label{fig:N1}}

The question of the impact on the value of $\hat{q}$
 of choosing $\hat{\omega}$ -
instead of $\omega_{BH}$ - as the cut-off for the spectrum,
 should now be discussed.
Indeed the above indicates that although 
larger than expected from leading order estimates based on the presence
of a thermalized and ideal QGP (Appendix B),
the values obtained above for $\hat{q}$ are much smaller than the ones quoted in 
\cite{Eskola:2004cr}, namely $5 < \hat{q} < 15~GeV^2/fm$.

An even smaller value may be obtained with $L = 5~fm$,
 namely $\hat{q} \simeq 0.7 ~GeV^2/fm$.
In a thermal gluonic system this implies 
an ($\underline{\rm{time-averaged}}$) 
 energy density of $\epsilon \simeq 4~GeV/fm^3$.

It is obvious
that the values of $\hat q$ and $L$ are strongly correlated, namely a large
transport coefficient, corresponding to a dense medium implies a
shorter path length $L$, and vice versa.
For a realistic average path length of 
$L \simeq 3~fm$ in the case of Au-Au collisions under consideration
the prefered value of the $\underline{\rm{time-averaged}}$ 
transport coefficient becomes
\begin{equation}
\hat{q} \simeq 2 ~GeV^2/fm~,
\label{eq:c2}
\end{equation}
which may still be accomodated into the pQCD
framework, at least within the uncertainties of LO approximations,
contrary to the "strong" QGP values of \cite{Eskola:2004cr}.

In fact, the actual values of the cut-off cannot be left out of the discussion.
Taking numbers quoted above:
$\hat{q} = 10~ GeV^2/fm$ and $L=2 ~fm$, leads to $\omega_c = 100 ~GeV$. If as 
indicated  above, we take $\hat{\omega}/\omega_c \simeq 10^{-2}$, one finds
$\hat{\omega} \simeq 1~GeV$,
 and thus correspondingly $\omega_{BH} \simeq 3-4~ GeV$, which is too large to 
make sense for energies/transverse momenta under consideration!
If, on the other hand, we want to have a reasonable value of $\omega_{BH} 
 \simeq 1.4~ GeV$, from the start, imposing correspondingly that the value of 
$\hat{\omega}$ is a factor $3-4$ smaller and keeping 
$\hat{\omega}/\omega_c \simeq 10^{-2}$, we find $\omega_c$ and thus
$\hat{q}$,   3 -4 times 
smaller.
We  use the value of $Q_{max} \le 0.3$  and $Q_{min} \simeq 0.2$
 as a way to constrain 
the relevant parameters: from $Q_{max}$, we take the cut-off to be 
$\simeq 1.4~ GeV$. 
This cut-off can only be  $\omega_{BH}$.  Then from $Q_{min} \simeq 0.2$, we deduce 
$\omega_c$. Finally, we find, depending on the length $L$, reasonably small values of 
$\hat{q}$.

A  determination of the average  $L$ should be possible with the expression for
$Q_{min} (p_\bot)$, (\ref{eq:e10}), which depends on $L$.
 Since
$\omega_{BH} \ll \omega_c$, we use to a good approximation 
 (\ref{eq:e5}) to determine
$N(\omega_{BH}) = N (\omega_{BH}/\omega_c)$
and insert (neglecting logarithmic factors)
\begin{equation}
\frac{2 \omega_c}{\omega_{BH}} \simeq \frac{L^2}{\lambda_g^2} = 
\left( \frac{N_c}{C_F} \right)^2 \left( \frac{L}{\lambda_q} \right)^2~,
\label{eq:PL2}
\end{equation}
defining the quark mean free path $\lambda_q = \frac{N_c}{C_F} \lambda_g$,
such that
\begin{equation}
Q_{\min} (p_\bot) \simeq \exp \left\{- \frac{ 2 \alpha_s C_F}{\pi}
\left[ \frac{N_c}{C_F} \frac{L}{\lambda_q} + \ln 2 \ln \left(
\frac{2 \lambda^2_q C^2_F}{L^2 N_c^2} \right) -1.44 \right] \right\}~.
\label{eq:PL3}
\end{equation}

In leading order $L/\lambda_q \gg 1$ the dependence with respect to
the path length has the typical characteristic behaviour of a survival
probability $\exp \left[ - L/\lambda_{eff}\right]$, where $\lambda_{eff}
\simeq \frac{\pi}{2 \alpha_s N_c} \lambda_q \simeq \lambda_q$ for 
$\alpha_s \simeq 1/2$. 
Without further geometrical restrictions the mean
path length $\langle L \rangle$ would be just given by the mean free path 
of a quark jet in
the medium, $\langle L \rangle \simeq \lambda_q$.

A better estimate of $ L / \lambda_q$, however, is obtained by
taking the full expression (\ref{eq:PL3}) into account:
for $Q_{min} \simeq 0.2$ the corresponding ratio is
 $ L / \lambda_q \simeq 3.35$.

In order to obtain $L$, we estimate the mean free path $\lambda_q =
9/4 \lambda_g$ from $\hat{q}$ and $\omega_{BH}$ by
\begin{equation}
\lambda_q \simeq \frac{9}{4} \sqrt{\frac{\omega_{BH}}{\hat{q}}}
\simeq 0.9~ fm ~,
\label{eq:PL3a}
\end{equation}
with $\hat{q} \simeq 2~GeV^2/fm$ and $\omega_{BH} \simeq 1.6~GeV$ and
find $L \simeq 3~fm$, consistent with the prefered value given above.
Note the related estimate for the mass $\mu$:  $~\mu \simeq 0.9 ~GeV$.

\subsection{Nuclear geometry}

So far we have considered averaged values for $Q_{min}, Q_{max}$
without taking the nuclear geometry explicitely into account.
We now present a more detailed discussion which, as we shall show, leads
on a firmer basis to similar conclusions as above as far as the medium
parameters are concerned.
We assume head-on collisions and essentially cylinder-like Au nuclei.
The quark jet is produced in Au-Au collisions at
mid-rapidity and propagates in the transverse plane.
Following \cite{Eskola:2004cr}, 
one starts from the geometrical transverse
path length
\begin{equation}
L_{\rm geom}{(\vec s)} = - s \cos \phi_{LS} + \sqrt{s^2 \cos^2 \phi_{LS}
+ R^2_{Au} - s^2},
\label{eq:PL4}
\end{equation}
where the position at which the parton is produced is denoted by the
vector $\vec s$ in the transverse plane. $\phi_{LS}$ is the angle of 
propagation with respect to this vector. This geometrical picture allows us
to obtain an average value for $\langle Q_{min} \rangle$ by calculating
\begin{equation}
\langle Q_{min}(\hat{q}/\omega_{BH} )\rangle =
 \frac{\int d^2s  \exp
\left\{ - N(L_{\rm geom})\right\} }{
\pi R^2_{Au}}~,
\label{eq:PL5}
\end{equation}
with $\vert \vec s \vert \le R_{Au}$, the radius of the Au nucleus.
In order to obtain $N(L_{\rm geom})$ we use (\ref{eq:e5}) with
 $\omega_c/\omega_{BH}
= \hat{q}/(2 \omega_{BH}) ~L^2_{\rm geom}$.

\FIGURE{\epsfig{file=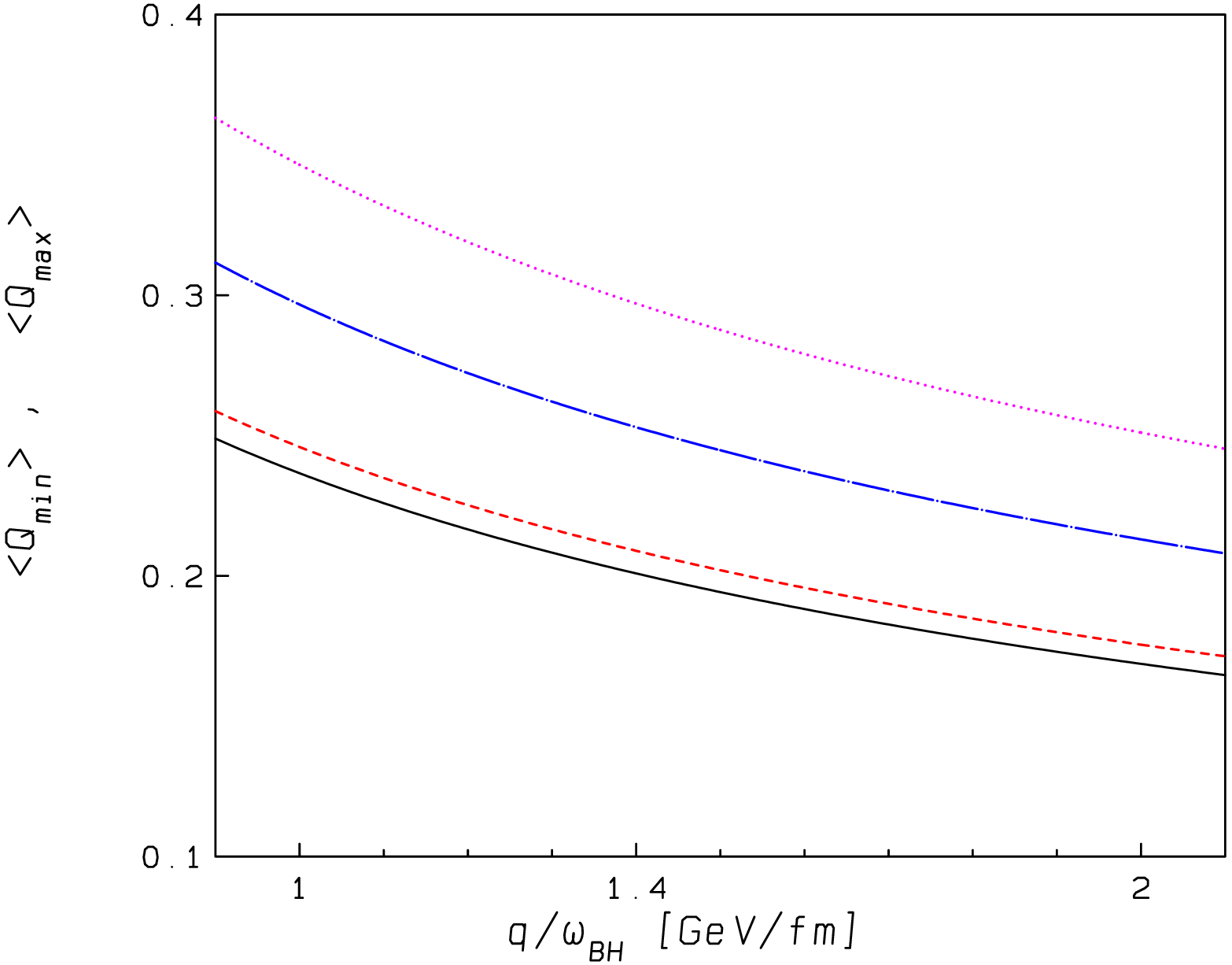, width=100mm}
 \caption{$\langle Q_{min} \rangle $ (solid curve)  and $\langle Q_{max} \rangle$,  respectively, as  functions of $\hat{q}/\omega_{BH}$
according to (\ref{eq:PL5}). $\langle Q_{max} \rangle$ for different values of $\omega_{BH}$:$0.75$ (dotted), $1.15$ (dashed-dotted) and $2.75 ~ GeV$ (dashed curve).}\label{fig:qmin1}}

In Fig.~\ref{fig:qmin1} we plot $\langle Q_{min} \rangle$
as a function of $\hat{q}/\omega_{BH}$, and observe that
$\langle Q_{min} \rangle \simeq 0.2$ for
\begin{equation}
\hat{q}/\omega_{BH} \simeq 1.4~ \frac{GeV}{fm}~.
\label{eq:PL5a}
\end{equation}
In the same Fig.~\ref{fig:qmin1} we plot $\langle Q_{max} \rangle$,
obtained analogously to (\ref{eq:PL5}),
for different values of $\omega_{BH}$ (and for $n/p_{\perp} = 1/GeV$).

In order to have  $\langle Q_{max} \rangle < 0.3$  - together with 
$\langle Q_{min} \rangle \simeq 0.2$ - we find 
$0.75 <  \omega_{BH} < 2.0 ~ GeV$.

In Fig.~\ref{fig:qlq1} we show $\langle Q_{min} \rangle $, but
 this time as a function
of $\lambda_q$, obtained from
\begin{equation}
\langle Q_{min}(\lambda_q)\rangle =
 \frac{\int d^2s ~ 
Q_{min} (L_{\rm geom}/\lambda_q)}{
\pi R^2_{Au}}~,
\label{eq:PL7}
\end{equation}
after inserting $L = L_{geom}$ in (\ref{eq:PL3}).
This way we find $\lambda_q \simeq 0.85~ fm ~~(\lambda_g \simeq 0.38~ fm)$,
when $Q_{min} \simeq 0.2$, in good agreement with (\ref{eq:PL3a}).

From  Fig.~\ref{fig:qlq1} we deduce that $0.65 <  \mu < 1.1 ~GeV$.
Finally, within these bounds the transport coefficient
$\hat{q} \simeq \mu^2/\lambda_g$ 
becomes
\begin{equation}
\hat{q} \simeq 1.0 - 3.0 ~ GeV^2/fm ~.
\label{eq:PL9}
\end{equation}
Averaging with respect to the nuclear geometry in this
straightforward manner leads to a final estimate which is compatible
with (\ref{eq:c2}). The "uncertainty" given by (\ref{eq:PL9})
 may be considered as
"theoretical error" on the derived medium parameters.

Concerning the robustness of the estimate  (4.10) for $\hat{q}$,
one has to keep in mind that
besides the IR sensitivity under discussion,  all the estimates
are based on LO QCD calculations.
Nevertheless, the explanation of quenching as being due to the LPM radiation
in a perturbative regime appears to be  a robust
statement.

\FIGURE{\epsfig{file=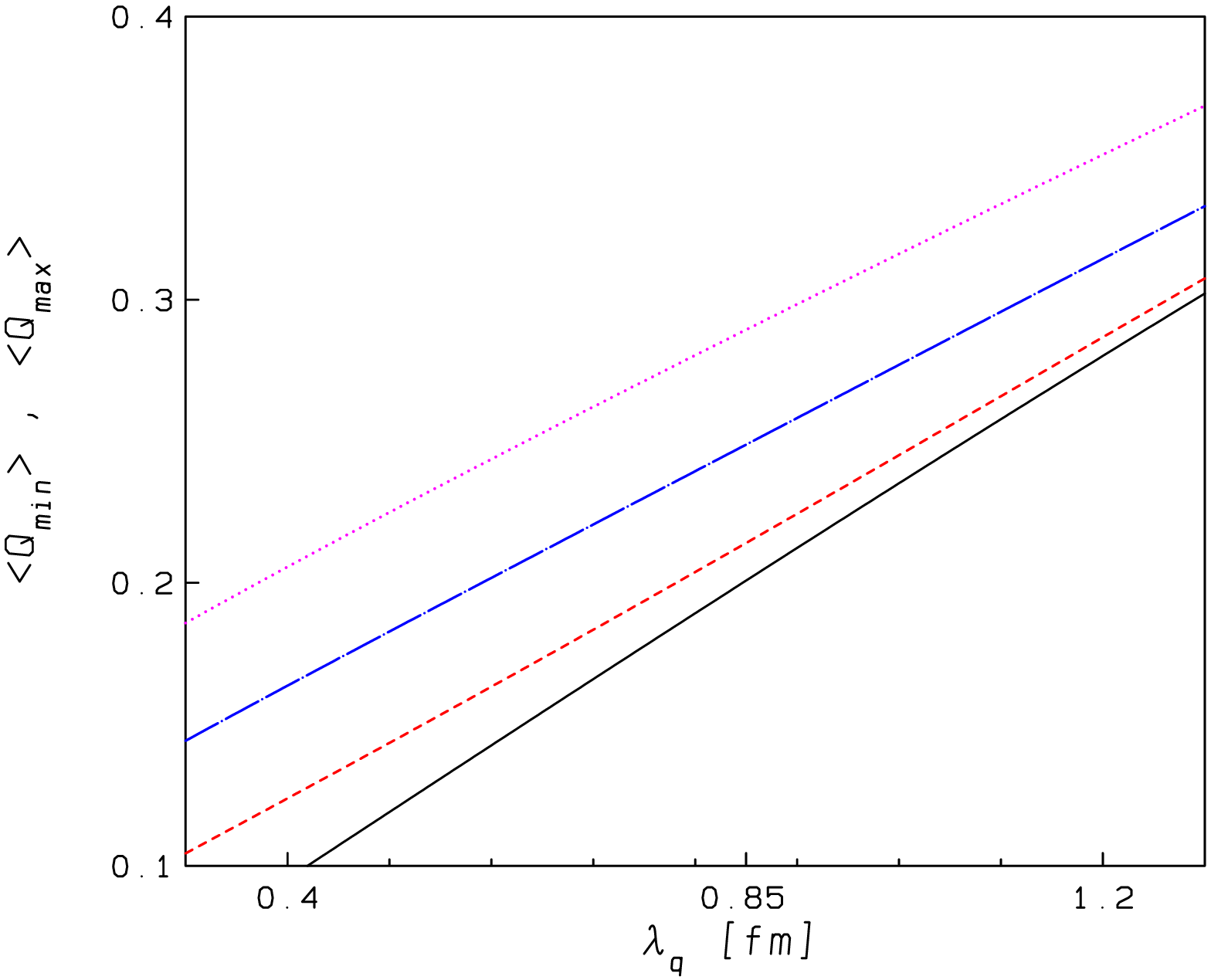, width=100mm}
\caption{$\langle Q_{min} \rangle$ (solid curve) and $\langle Q_{max} \rangle$,
 respectively,  as functions
 of $\lambda_{q}$
according to (\ref{eq:PL7}). 
$\langle Q_{max} \rangle$
for different values of the screening mass $\mu$:
$0.65$ (dotted), $0.8$ (dashed-dotted) and $1.1~GeV$ (dashed curve).}\label{fig:qlq1}}

A value of $\hat{q} \simeq 1.8 ~GeV^2/fm$ would correspond to
a temperature of $T \simeq 375~MeV$, corresponding to an
average energy density of $\epsilon \simeq 12.5 ~GeV/fm^3$,
in agreement with the results quoted 
in \cite{Turbide:2005fk}.

In order to obtain the actual value of $\hat{q}$
at the very early stage of the collisions, at times
$\tau \simeq 1/p_{\perp} < 2 \cdot 10^{-2}~fm$, one has to include
the effects due to the longitudinal \cite{Baier:1998yf,Salgado:2002cd},
 but also transverse \cite{Renk:2005ta}
 expansion of the dense system during the time of $O(L)$,
the jet takes to propagate through this medium.

\acknowledgments

We would like to thank  F.~Arleo, Yu.~L.~Dokshitzer, F.~Gelis, A.~H.~Mueller,
S.~Peign\'e, C.~Salgado and U.~A.~Wiedemann for stimulating discussions
and useful comments. Thanks are due to M. Van Leeuwen for helpful criticisms.
R.~B. is grateful for the kind hospitality extended to him at LPT, Orsay,
where most of this work has been done.

\appendix
\section{Estimate of the Bethe-Heitler and absorption contributions}

In the plausible context, where the medium is thermalized, let us investigate 
how the contributions of the Bethe-Heitler and the absorption spectra
modify the analysis discussed so far in the paper.

In LO pQCD the Bethe-Heitler-Gunion-Bertsch \cite{Gunion} differential spectrum
for inclusive gluon production in a medium of length $L$ 
in the presence of $L/{\lambda_g}$  scatterers reads
\begin{equation}
\frac{d I}{dy d^2 p_\bot} = \frac{\alpha_s C_F}{\pi^2} \frac{1}
{p^2_\bot} \left( \frac{L}{\lambda_g} \right) , 
\label{(A.1)}
\end{equation}
which after $p_\bot$-integration becomes
\begin{equation}
\omega \frac{d I}{d\omega}\bigg|_{BH} \simeq \frac{\alpha_s C_F}{\pi} \ln
\frac{\omega^2_{BH}}{\omega^2_{{\rm cut}}} \left( \frac{L}{\lambda_g} 
\right)~.
\label{(A.2)}
\end{equation}
 Because of the presence of the 
(non-perturbative) IR-cut, $\omega_{{\rm cut}}$, this BH-intensity is not 
precisely determined. Nevertheless, an estimate of the gluon energy 
$\omega_{BH}$ may be obtained from matching
(\ref{(A.2)}) with the LPM expression for the 
intensity at $\omega = \omega_{BH}$,
\begin{equation}
\omega \frac{dI}{d \omega} \bigg|_{LPM} \simeq \frac{\alpha_s C_F}{\pi}
\sqrt{\frac{2 \omega_c}{\omega_{BH}}} , 
\label{(A.3)}
\end{equation}
valid for $\omega \ll \omega_c$. When logarithmic factors are 
taken to be of $O(1)$, i.e. $\omega_{BH} \simeq 1.65~ \omega_{\rm{cut}}$,
we find
\begin{equation}
\omega_{BH} \simeq \mu^2 \lambda_g , 
\label{(A.4)}
\end{equation}
in terms of the screening mass $\mu$ and the mean free path of the gluon
$\lambda_g$.

The resulting quenching factor becomes
\begin{equation}
Q_{BH} (p_\bot ) =  \exp \left\{ - \int^{\omega_{BH}}_{\omega_{{\rm cut}}}
~ \frac{dI}{d \omega}\bigg|_{BH}  
\left[ 1 - e^{n \omega /p_\bot} \right] \right\} ~,
\label{(A.5)}
\end{equation}
and expanding the integrand in the limit of small $\omega$, we find
with (\ref{(A.2)})
\begin{equation}
Q_{BH} (p_\bot ) \ge 
 \exp \left\{ - \frac{\alpha_s C_F}{\pi}
 \left( \frac{L}{\lambda_g} \right) 
 \frac{n}{p_\bot} (\omega_{BH} - \omega_{{\rm cut}}) 
\right\} ~.
\label{(A.44)}
\end{equation}

Following \cite{Turbide:2005fk}, we 
consider the absorption contribution in the
presence of a heat bath. It is assumed that for the radiation energy
$\omega < 0$ the spectrum is approximated by 
\begin{equation}
\frac{dI}{d\omega} \bigg|_{abs} \simeq \frac{\alpha_s C_F}{\pi} 
\frac{1}{|\omega |} \left( \frac{L}{\lambda_g} \right) 
e^{- |\omega | / T} , 
\label{(A.6)}
\end{equation}
i.e. the Bethe-Heitler spectrum multiplied by a Boltzmann factor
with temperature $T$. 
The corresponding quenching factor $Q_{abs} (p_\bot)$ then becomes
\begin{eqnarray}
Q_{abs} (p_\bot ) &=& \exp \left\{ - \int^\infty_0 \frac{dI}{d | \omega |}
\left[ 1 - e^{n |\omega |/p_\bot} \right] \right\} \nonumber \\
&= & \exp \left\{ - \frac{\alpha_s C_F}{\pi} \frac{L}{\lambda_g} 
\int^\infty_0 \frac{d\omega}{\omega} e^{- \omega/T} \left[ 1 - e^
{n\omega / p_\bot} \right] \right\} . 
\label{(A.7)}
\end{eqnarray}
The integral may be approximated by 
\begin{equation}
\int^\infty_0 \frac{d \omega}{\omega} e^{-\omega/T} \left[1 - e^{n \omega 
/ p_\bot} \right] \simeq - \frac{n}{p_\bot} \int^\infty_0 d\omega e^{
-\omega / T} = - \frac{nT}{p_\bot} , 
\label{(A.8)}
\end{equation}
leading to 
\begin{equation}
Q_{abs} (p_\bot ) \ge \exp \left\{ \frac{\alpha_s C_F}{\pi} \left(
\frac{L}{\lambda_g} \right) \frac{nT}{p_\bot} \right\} . 
\label{(A.9)}
\end{equation}

Finally, multiplying the two quenching factors $Q_{BH} (p_\bot )$ 
and $Q_{abs} (p_\bot)$, leads to a lower bound 
\begin{equation}
 Q_{BH}(p_{\perp}) Q_{abs}(p_{\perp})
 \ge \exp \left\{ + \frac{\alpha_s C_F}{\pi}
 \left( \frac{L}{\lambda_g} \right) 
 \frac{n}{p_\bot} \left[ T- (\omega_{BH} - \omega_{{\rm cut}})  \right] 
\right\} ~.
\label{(A.10)}
\end{equation}
Taking as typical numbers: $\omega_{BH} \simeq 1.6~GeV$, $\omega_{{\rm cut}}
\simeq 1 ~GeV$, $T = 350 ~MeV$, $p_{\perp}= 10~GeV,~  n = 10$, and
$L = 3 ~fm, ~ \lambda_g = 0.38 ~fm$, we find 
\begin{equation}
 Q_{BH} (p_\bot) Q_{abs} (p_\bot ) \ge 0.7~.  
\label{(A.12)}
\end{equation}

This indicates that we may as a first guess, as already suggested in
\cite{Turbide:2005fk}, neglect altogether the Bethe-Heitler
 and absorption processes, meaning that the values we thus obtain for
$\hat{q}$ are in fact upper bounds, and therefore comforting
the perturbative framework.

\section{Thermal parameters}

Let us consider  an  equilibrated system $(N_c = 3, N_f = 0)$
in the weak coupling QCD limit \cite{LeBellac}
 and give a   
short summary of the  elements which enter our analysis.

 In LO given the temperature $T$ the screening mass is 
\begin{equation}
\mu^2 = 4 \pi \alpha_s T^2 . 
\label{(B.1)}
\end{equation}
The gluon mean free path $\lambda_g$ is expressed in terms of the 
gluon density
\begin{equation}
\rho_g = \frac{16}{\pi^2} \zeta (3) T^3~ ,~~~ \zeta (3) = 1.202~, 
\label{(B.2)}
\end{equation}
and the (transport) gluon-gluon cross section (to logarithmic accuracy)
\begin{equation}
\sigma^{gg}_T \simeq \frac{N_c}{C_F} \frac{2 \pi \alpha^2_s}{ \mu^2}
 \ln \left( 1 / \alpha_s \right) \simeq \frac{9 \pi \alpha^2_s}{2 \mu^2} ,
\label{(B.3)}
\end{equation}
when neglecting logarithmic dependence, i.e. 
\begin{equation}
1 / \lambda_g = \rho_g \sigma^{gg}_T \simeq \frac{18}{\pi^2} \zeta (3) 
\alpha_s T  \simeq 2.2 ~\alpha_s T~. 
\label{(B.4)}
\end{equation}
The mean free path for a quark is $\lambda_q = 9/4 \lambda_g$.
The corresponding energy density of this gluonic system is 
\begin{equation}
\epsilon = \frac{8 \pi^2}{15} T^4 . 
\label{(B.5)}
\end{equation}

Typical orders of magnitude may be given e.g for a temperature of 
$T = 400~ MeV$ and a coupling $\alpha_s = 1/2$: the screening mass is 
$\mu \simeq 1~ GeV$, the mean free path
 $\lambda_g \simeq 0.45~ fm ~~(\lambda_q 
\simeq 1~ fm)$, implying $\omega_{BH} \simeq \mu^2 \lambda_g \simeq 
2.25~ GeV$. The transport coefficient is 
estimated as $\hat q \simeq \mu^2 / \lambda_g 
\simeq 2.2 ~ GeV^2 / fm$, 
leading to an energy density of $\epsilon \simeq 17~  GeV/ fm^3$
when using
 $\hat{q}~\simeq 2~ \epsilon ^{3/4}$ \cite{Baier:2002tc}.

\end{document}